\documentclass[dvips]{article}
\usepackage{icrctc07}

\title{HESS~J1023--575: Non-thermal particle acceleration associated \\ with the young stellar cluster Westerlund~2}
\shorttitle{VHE gamma-rays from Westerlund~2}
\authors{O.~Reimer$^{1}$, J.~Hinton$^{2}$, W.~Hofmann$^{3}$, S.~Hoppe$^{3}$, C.~Masterson$^{4}$, M.~Raue$^{5}$,\\
 for the H.E.S.S. Collaboration$^{6}$.}
\shortauthors{O. Reimer et al}
\afiliations{
$^1$ Stanford University, HEPL \& KIPAC, Stanford, CA 94305-4085, USA\\ 
$^2$ School of Physics \& Astronomy, University of Leeds, Leeds LS2 9JT, UK\\
$^3$ Max-Planck-Institut f\"ur Kernphysik, P.O. Box 103980, 69029 Heidelberg, Germany\\
$^4$ Dublin Institute for Advanced Studies, 5 Merrion Square, Dublin 2, Ireland\\
$^5$ Universit\"at Hamburg, Institut f\"ur Experimentalphysik, Luruper Chaussee 149, 22761 Hamburg, Germany\\
$^6$ http://www.mpi-hd.mpg.de/hfm/HESS/public/hn\_hesscollab.html
}
\email{olr@stanford.edu; martin.raue@desy.de}

\abstract{The results from H.E.S.S. observations towards Westerlund~2 are presented. 
The detection of very-high-energy gamma-ray emission towards the young stellar cluster
Westerlund 2 in the HII complex RCW49 by H.E.S.S. provides ample evidence that
particle acceleration to extreme energies is associated with this region. A 
variety of possible emission scenarios is mentioned, ranging from high-energy 
gamma-ray production in the colliding wind zone of the massive Wolf-Rayet 
binary WR~20a, collective wind scenarios, diffusive shock acceleration at the 
boundaries of wind-blown bubbles in the stellar cluster, and outbreak phenomena 
from hot stellar winds into the interstellar medium. These scenarios are briefly 
compared to the characteristics of the associated new VHE gamma-ray source HESS~J1023--575, 
and conclusions on the validity of the respective emission scenarios for 
high-energy gamma-ray production in the Westerlund~2 system are drawn.}

\begin{document}
\maketitle

\section{The young stellar cluster Westerlund~2 in the HII region RCW~49}
The prominent giant HII region RCW 49, and its ionizing young stellar cluster Westerlund 2, are located 
towards the outer edge of the Carina arm of our Milky Way. RCW 49 is a luminous, massive star formation region, 
and has been extensively studied at various wavelengths. Recent mid-infrared measurements with SPITZER 
revealed still ongoing massive star formation \cite{ref1}. The regions surrounding Westerlund 2 appear evacuated 
by stellar winds and radiation, and dust is distributed in fine filaments, knots, pillars, bubbles, 
and bow shocks throughout the rest of the HII complex \cite{ref2, ref3}. Radio continuum observations revealed 
two wind-blown shells in the core of RCW~49 \cite{ref4}, surrounding the central region of Westerlund~2, 
and the prominent Wolf-Rayet star WR~20b. A long-standing distance ambiguity has been recently \cite{ref5} 
revised in a determination of the distance to Westerlund~2 by spectro-photometric measurements 
of 12 cluster member O-type stars of $(8.3 \pm 1.6)$ kpc. This value is in good agreement with the 
measurements from the light curve of the eclipsing binary WR~20a \cite{ref6}, associating WR 20a as a cluster 
member of Westerlund 2 (Note, however the 2.8 kpc as of \cite{ref_fn}). The stellar cluster contains an extraordinary ensemble of hot and massive stars, 
presumably at least a dozen early-type O-stars, and two remarkable WR stars. Only recently WR20a was 
established to be a binary \cite{ref7, ref8} by presenting a solutions for a circular orbit with a period of 3.675, and 3.686 days, 
respectively. Based on the orbital period, the minimum masses have been found to be $(83 \pm 5)$\,M$_{\odot}$ and $(82 \pm 5)$\,M$_{\odot}$ 
for the binary components \cite{ref6}. At that time, it classified the WR binary WR 20a as the most massive of all confidently 
measured binary systems in our Galaxy. The supersonic stellar winds of both WR stars collide, and a wind-wind 
interaction zone forms at the stagnation point with a reverse and forward shock. In a detached binary system 
like WR~20a, the colliding wind zone lies between the two stars, and is heavily skewed by Coriolis forces. 
The winds of WR~20a can only be accelerated to a fraction of their expected wind speed $v_\infty\sim 2800$ km/s, and a 
comparatively low pre-shock wind velocity of $\sim 500$ km/s follows. Synchrotron emission has not yet been detected 
from the WR~20a system, presumably because of free-free-absorption in the optically thick stellar winds along 
the line of sight. WR~20a has been detected in X-rays \cite{ref9}, but non-thermal and thermal components of the X-ray 
emission remain currently indistinguishable. Detectable VHE gamma-radiation from the WR~20a binary system was 
only predicted in a pair cascade model \cite{ref10}, although detailed modeling of the WR~20a system in other scenarios 
(e.g. as of \cite{ref19} when produced either by optically-thin inverse Compton scattering of relativistic electrons with the dense photospheric 
stellar radiation fields in the wind-wind collision zone or in neutral pion decays, with the mesons produced by inelastic 
interactions of relativistic nucleons with the wind material) is still pending. 
At VHE gamma-rays, photon-photon absorption will diminish the observable flux from a close binary system such as WR~20a 
\cite{ref11}.

\section{H.E.S.S. observations of Westerlund~2}

The H.E.S.S. (High Energy Stereoscopic System) collaboration observed the Westerlund~2 region between March and July 2006, 
and obtained 14 h (12.9 h live time) of data, either on the nominal source location of WR~20a or overlapping data from the 
ongoing Galactic plane survey. Standard quality selections were imposed on the data. The data have been obtained in 
wobble-mode observations to allow for simultaneous background estimation. The wobble offsets for these observations 
range from 0.5$^\circ$ to 2$^\circ$, with the majority of data taken with wobble offset less than 0.8$^\circ$. The zenith angles range 
between 36$^\circ$ and 53$^\circ$, resulting in an energy threshold of 380 GeV for the analysis. The data have been analyzed using 
the H.E.S.S. standard Hillas analysis with standard cuts ($>$ 80 p.e.). A point source analysis on the nominal position 
of WR~20a resulted in a clear signal with a significance of 6.8$\sigma$. Further investigations revealed an extended excess 
with a peak significance exceeding 9$\sigma$ (Fig.3 left). The center of the excess was derived by fitting the two-dimensional 
point spread function (PSF) of the instrument folded with a Gaussian to the uncorrelated excess map:  
$\alpha_{2000}$ = $10^{\rm h}23^{\rm m}18^{\rm s} \pm 12^{\rm s}$, $\delta_{2000}$ = -57$^\circ$45'50'' $\pm$ 1'30''. 
The systematic error in the source location is 20'' in both coordinates. The source is clearly extended beyond the nominal 
extension of the PSF (Fig.~1). A fit of a Gaussian folded with the PSF of the H.E.S.S. instruments gives an extension of $0.18^\circ \pm 0.02^\circ$. 

\begin{figure}[ht]
   \centering
   \includegraphics[width=0.48\textwidth]{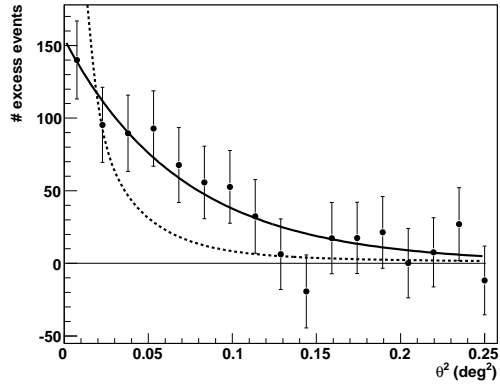} 
   \caption{Number of excess events versus the squared angular distance from the best fit position of the excess. 
            The dashed line shows the expectation for a point source derived from Monte Carlo data. The solid line is a 
			fit of the PSF folded with a Gaussian ($\sigma = 0.18^\circ \pm 0.02^\circ$).}
   \label{fig1}
\end{figure}

The differential energy spectrum for photons inside the corresponding 85\% containment radius of 0.39$^\circ$ is shown in Fig.~2. 
It can be described by a power law (dN/dE$= \Phi_0 \cdot (\mbox{E}/1\,\mbox{TeV})^{- \Gamma}$) 
with a photon index of $\Gamma=2.53 \pm 0.16_{\mathrm{stat}} \pm 0.1_{\mathrm{syst}}$ and a normalization at 1\,TeV of 
$\Phi_0 = (4.50 \pm 0.56_{\mathrm{stat}} \pm 0.90_{\mathrm{syst}}) \times 10^{-12}$\,TeV$^{-1}$\,cm$^{-2}$\,s$^{-1}$. 
The integral flux for the whole excess above the energy threshold of 380 GeV is (1.3 $\pm$ 0.3) $\times 10^{-11}$\,cm$^{-2}$\,s$^{-1}$. 
No significant flux variability could be detected in the data set. The fit of a constant function to the lightcurve binned 
in data segments of 28\,minutes has a chance probability of 0.14. The results were checked with independent analyses and 
found to be in good agreement.

\begin{figure}[ht]
   \centering
   \includegraphics[width=0.48\textwidth]{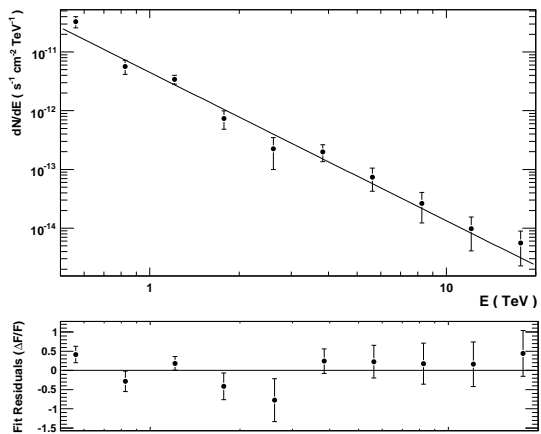}
   \caption{Differential energy spectrum and residuals to a single power-law fit of HESS~J1023--575 from photons 
            inside the 85\% containment radius (0.39$^\circ$) around the best fit position. The background is estimated with 
			background regions of the same size and distance from the camera center as the signal region.} 
      \label{fig2}
\end{figure}

\section{HESS~J1023--575 in the context of $\gamma$-ray emission scenarios}

The detection of VHE gamma-ray emission from the Westerlund~2 region \cite{ref12} is proof for extreme high-energy particle 
acceleration associated with this young star forming region. With a projected angular size of submilliarcsecond 
scale, the WR~20a binary system, including its colliding wind zone, would appear as a point source for observations 
with the H.E.S.S. telescope array. Unless there are extreme differences in the spatial extent of the particle 
distributions producing radio, X-ray, and VHE gamma-ray emission, a \emph{colliding stellar wind scenario} for the WR~20a
binary faces the severe problem of accounting for the observed VHE source extension. At a nominal distance of 8.0~kpc, this 
source extension is equivalent to a diameter of 28~pc for the emission region, consistent in size with theoretical 
predictions of bubbles blown from massive stars into the ISM \cite{ref13}. The spatial extension found for HESS~J1023--575 
contradicts emission scenarios where the bulk of the gamma-rays are produced close to the massive stars. Alternatively, 
the emission could arise from \emph{collective effects of stellar winds} in the Westerlund~2 stellar cluster. Diffusive shock 
acceleration in cases where energetic particles experience multiple shocks can be considered for Westerlund~2. 
The stellar winds may provide a sufficiently dense target for high-energy particles, allowing the production of 
$\pi^0$-decay $\gamma$-rays via inelastic pp-interactions. Collective wind scenarios \cite{ref14, ref15} suggest that 
the spatial extent of the gamma-ray emission corresponds to the volume filled by the hot, shocked stellar winds, 
but HESS~J1023--575 substantially supersedes the boundary of Westerlund~2. Supershells, molecular clouds, and inhomogeneities embedded 
in the dense hot medium may serve as the targets for gamma-ray production in Cosmic Ray interactions. Such environments 
have been studied in the nonlinear theory of particle acceleration by large-scale MHD turbulence \cite{ref16}. \emph{Shocks and MHD 
turbulent motion inside a stellar bubble or superbubble} can efficiently transfer energy to cosmic rays if the particle acceleration time 
inside the hot bubble is much shorter than the bubble's expansion time. 

\begin{figure*}
\begin{center}
\includegraphics [width=0.90\textwidth]{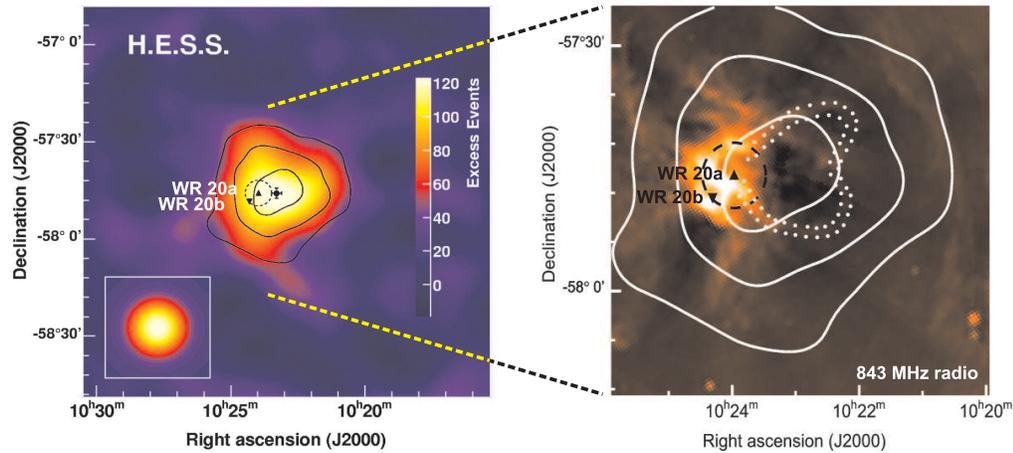}
\end{center}
\caption{
Left: H.E.S.S. $\gamma$-ray sky map of the Westerlund~2 region, smoothed to reduce the effect of statistical fluctuations. 
The inlay in the lower left corner shows how a point-like source would have been seen by H.E.S.S. WR~20a and WR~20b 
are marked as filled triangles, and the stellar cluster Westerlund~2 is represented by a dashed circle.    
Right: Significance contours of the $\gamma$-ray source HESS~J1023--575 (corresponding 5, 7 and 9$\sigma$ ), overlaid on a 
MOST radio image. The wind-blown bubble around WR~20a, and the blister to the west of it can be seen as depressions in 
the radio continuum. The blister is indicated by white dots as in \cite{ref4}, 
and appears to be compatible in direction and location with HESS~J1023--575.
}\label{fig3}
\end{figure*}

Finally, shock acceleration at the boundaries of the "blister" (Fig.~3 right) may enable particles to diffusively 
re-enter into the dense medium, thereby interacting in hadronic collisions and producing gamma-rays. A scenario as 
outlined in \cite{ref17} for a Supernova-driven expansion of particles into a low density medium may be applicable to the 
expanding stellar winds into the ambient medium. If one accepts such a scenario here, it might give the first 
observational support of gamma-ray emission due to diffusive shock acceleration from supersonic winds in a 
wind-blown bubble around WR~20a, or the ensemble of hot and massive OB stars from a superbubble in Westerlund~2, 
breaking out beyond the edge of a molecular cloud. Accordingly, one has to consider that such acceleration sites 
will also contribute to the observed flux of cosmic rays in our Galaxy \cite{ref18}.

Further observations with the H.E.S.S. telescope array will help to discriminate among the alternatives in the 
interpretation of HESS~J1023--575. However, the convincing association with a new type of astronomical object 
a massive HII region and its ionizing young stellar cluster profoundly distinguishes this new detection by 
the H.E.S.S. telescope array already from other source findings made during earlier Galactic Plane Scan observations.

\section{Acknowledgements}
\scriptsize
The support of the Namibian authorities and of the University of Namibia
in facilitating the construction and operation of H.E.S.S. is gratefully
acknowledged, as is the support by the German Ministry for Education and
Research (BMBF), the Max Planck Society, the French Ministry for Research,
the CNRS-IN2P3 and the Astroparticle Interdisciplinary Programme of the
CNRS, the U.K. Science and Technology Facilities Council (STFC),
the IPNP of the Charles University, the Polish Ministry of Science and 
Higher Education, the South African Department of
Science and Technology and National Research Foundation, and by the
University of Namibia. We appreciate the excellent work of the technical
support staff in Berlin, Durham, Hamburg, Heidelberg, Palaiseau, Paris,
Saclay, and in Namibia in the construction and operation of the
equipment.

\normalsize
\bibliographystyle{plain}
{}

\end{document}